\documentclass{aastex}
\usepackage{spr-astr-addons}
\usepackage{url}

\RequirePackage{color}

\begin{document}

\title{Interacting Kasner-type cosmologies}
\slugcomment{Not to appear in Nonlearned J., 45.}
%% Running heads
\shorttitle{Short article title}
\shortauthors{Autors et al.}

\author{Mauricio Cataldo\altaffilmark{1}}
\affil{Departamento de F\'\i sica, Facultad de Ciencias, Universidad
del B\'\i o--B\'\i o, Avenida Collao 1202, Casilla 5-C,
Concepci\'on, Chile.\\} \and
\author{Fabiola Ar\'evalo\altaffilmark{2}}
\affil{Departamento de F\'{\i}sica, Universidad de Concepci\'{o}n,\\
Casilla 160-C, Concepci\'{o}n, Chile.}

\and

\author{Patricio Mella\altaffilmark{3}}
\affil{Departamento de F\'{\i}sica, Universidad de Concepci\'{o}n,\\
Casilla 160-C, Concepci\'{o}n, Chile.}

\altaffiltext{1}{mcataldo@ubiobio.cl}
\altaffiltext{2}{farevalo@udec.cl}
\altaffiltext{3}{patriciomella@udec.cl}

\begin{abstract}
It is well known that Kasner-type cosmologies provide a useful
framework for analyzing the three-dimensional anisotropic expansion
because of the simplification of the anisotropic dynamics. In this
paper relativistic multi-fluid Kasner-type scenarios are studied. We
first consider the general case of a superposition of two ideal
cosmic fluids, as well as the particular cases of non-interacting
and interacting ones, by introducing a phenomenological coupling
function $q(t)$. For two-fluid cosmological scenarios there exist
only cosmological scaling solutions, while for three-fluid
configurations there exist not only cosmological scaling ones, but
also more general solutions. In the case of triply interacting
cosmic fluids we can have energy transfer from two fluids to a third
one, or energy transfer from one cosmic fluid to the other two. It
is shown that by requiring the positivity of energy densities there
always is a matter component which violates the dominant energy
condition in this kind of anisotropic cosmological scenarios.
\end{abstract}


%\section*{}
%\label{sec:intro}

\section{Introduction}
It is well known that the present cosmological observations allow
the theoretical cosmology, based on General Relativity, to conclude
that nearly $96 \%$ of the matter content in the Universe is of
types which have not been seen in the laboratory. Even more, there
is also indirect evidence suggesting that nearly $74 \%$ of the
matter content present in the Universe exerts a negative pressure
(dark energy). So one of the key problems of the current cosmology
is deciding which kinds of matter sources were and are present in
the Universe. This notable uncertainty about the kind of matter
filling the Universe, which is unique when we attempt to apply the
laws of physics to the evolution of the Universe, requires the
cosmologists to proceed in a multi-faceted manner
\cite{Padmanabhan}.

Usually cosmological models are constructed
under the assumption that the matter source is an idealized perfect
fluid. This assumption may be a good approximation to the matter
content of the Universe, however at earlier epochs there may not be
so negligible the effects of anisotropic matter fields such as for
example magnetic and electric fields, populations of collisionless
particles such as gravitons, photons or relativistic neutrinos;
long-wavelength gravitational waves, topological defects such as
cosmic string and domain walls, among others \cite{Barrow}.

On the other hand, in recent years the current Cosmology has been
strongly influenced by the huge improvement in quality, quantity and
the scope of cosmological observations
\cite{Obs3,Obs7,Obs6,Obs4,Obs2,Obs5, Obs1}. Recent investigations
detect anisotropy in the cosmic microwave background radiation
(CMBR). The temperature anisotropy in the CMBR has been arguably the
most influential of these recent cosmological observations and
represents one of the most accurate observational data in the modern
cosmology. This scenario leads cosmologists to consider for early
stages of the Universe not only the standard isotropic and
homogeneous FRW metrics but also to consider more general universe
models. In principle one can consider inhomogeneous models which, in
order to reproduce the late homogenous and isotropic stages of the
Universe, must contain a FRW background for certain limits or values
of the model functions or parameters respectively. On the other hand
one can also consider anisotropic backgrounds. The simplest
generalization of FRW spacetimes are the homogeneous and anisotropic
Bianchi cosmologies of types IX, V and I, which are generalizations
of closed, open and flat FRW spacetimes respectively.

In the latter Bianchi type I we find the special class of
Kasner-type geometries, for which cosmological scale factors evolve
as a power law in time. It is important to notice that Kasner-type
geometries themselves have been used as a simple arena for
discussing some properties of anisotropically expanding cosmologies.
It is very natural for anisotropic Kasner-type cosmologies to
introduce a matter source described by an imperfect fluid with a
stress-energy tensor containing the coefficients of bulk viscosity,
shear viscosity and heat conduction. In
Refs.\cite{Brevik,Cataldo151,Cataldo152} are discussed anisotropic
matter sources for this kind of cosmologies. It was shown that in
general relativity the Bianchi type I metric of the Kasner form is
not able to describe an anisotropic universe filled with a viscous
fluid, satisfying simultaneously the dominant energy condition (DEC)
and the second law of thermodynamics \cite{Cataldo151}. In
Ref.\cite{ScTT} it is proved that this is possible in scalar tensor
theories, while for a more general Bianchi type I metric with a
perfect fluid it is possible in the Brans-Dicke theory of gravity
\cite{BD}.

On the other hand, in Ref.\cite{HOLOGRAFIA} the discussion goes into
the framework of the holographic principle, in Ref.\cite{Halpern}
the behavior of Kasner-type Cosmologies with Induced Matter is
studied, while in other more general and interesting contexts Kasner
geometries also are considered
\cite{Others2,Others5,Others1,Others6,Others4,Others3}. Lastly, let
us note that the Kasner metric plays an important role also in
Bianchi cosmologies of type IX (the Mixmaster Universe), where in
vacuum solutions the Mixmaster universe has served as a theoretical
playground for many ideas related to the question of the nature of
the chaotic behavior exhibited in some solutions of the vacuum
Einstein equations \cite{Bini}, or for models containing matter
where the spatial curvature causes the axes and rates of contraction
to undergo sudden jumps from one Kasner-like solution to another
\cite{Turok}. The Kasner spacetimes are often used for the
description of the very early stages of the Universe.

In this paper we shall consider Kasner-type cosmologies dominated by
two or three matter components.  The interaction among ideal fluid
components also will be considered. The interest in interacting
cosmologies has been focused mainly on FRW cosmologies in order to
address the observed late acceleration of the Universe
\cite{Armendariz1,Armendariz2,Armendariz3,Armendariz4} and the so
called cosmological coincidence problem
\cite{Cataldos1,Cataldos2,Cataldos3,Cataldos4}. Usually the universe
is modeled with perfect fluids and with mixtures of non-interacting
perfect fluids. This means that it is assumed that there is no
conversion (energy transfer) among the components and that each of
them evolves separately according to standard conservation laws.
However, we can consider plausible cosmological models containing
fluids which interact with each other, so the energy from one of the
fluids is diluted or decayed into another fluid component. The
advantage of considering multi-fluid components in anisotropic
Kasner-type cosmologies is that the problem is solved exactly, in an
analytical manner.

The outline of the present paper is as follows: In Section II we
establish and solve the Einstein field equations for two-fluid
Kasner-type cosmologies. In Section III solutions for three-fluid
configurations are considered. Finally, Section IV presents some
concluding remarks.

\section{Einstein field equations for two-fluid Kasner-type cosmologies}
We shall consider in this paper anisotropic cosmologies described by
the metric
\begin{equation}\label{KM}
ds^2=dt^2-t^{2p_1}dx^2-t^{2p_2}dy^2-t^{2p_3}dz^2,
\end{equation}
where we shall call $p_1$, $p_2$ and $p_3$ Kasner parameters and
they are constant ones. We shall refer to the Kasner-type
cosmologies (\ref{KM}) as simply Kasner cosmologies.

Thus, the Einstein´s field equations $R_{\mu \nu}-\frac{1}{2}Rg_{\mu
\nu}=-\kappa T_{\mu \nu}$ reduce to
\begin{eqnarray}
\label{ro}  \kappa \rho &=& \frac{p_1 p_2 + p_1 p_3 + p_2 p_3}{t^2},
\end{eqnarray}
\begin{eqnarray}
\label{P1} \kappa P_x &=& -\frac{p_2^2 + p_3^2 -p_2 -
p_3 + p_2 p_3}{t^2}, \\
\label{P2}  \kappa P_y &=& -\frac{p_1^2 + p_3^2 -p_1 - p_3 + p_1
p_3}{t^2}, \\
\label{P3}  \kappa P_z &=& -\frac{p_1^2 + p_2^2 -p_1 - p_2 + p_1
p_2}{t^2}.
\end{eqnarray}
Here $\kappa=8 \pi G$ and $P_j$, with $j = x,y,z$, representing the
effective momenta in the corresponding coordinate axis, implying
that in general for a Kasner metric we have anisotropic pressures.
In other words, for the metric (\ref{KM}) one might introduce a
matter source described by an imperfect fluid. Thus for a Kasner
space–-time one might consider a stress-energy tensor of an
imperfect fluid containing the coefficients of bulk viscosity, shear
viscosity and heat conduction. Note that in this case the energy
density and all pressures scale as $1/t^2$, implying that we have
barotropic equations of state for all pressures, i.e. $P_j=\omega_j
\rho$, with $\omega_j$ constants. However, in this paper we shall
consider only perfect cosmic fluids, i.e. $P_x=P_y=P_z$. Note that
in this case either $\rho$ and $P_j $ (with $j=x,y,z$) scale always
as $t^{-2}$.

Let us now consider a Kasner cosmology filled with two ideal cosmic
fluids $\rho_1$ and $\rho_2$. Thus by putting into
Eqs.(\ref{ro})--(\ref{P3}) the expressions $\rho=\rho_1+\rho_2$ and
$P_x=P_y=P_z=P_1+P_2$ we may rewrite the field equations in the
following form:
\begin{eqnarray}
S-Q &=&\frac{1}{2} \kappa t^2 \left(\rho_1+\rho_2+3(P_1+P_2)
\right),\label{1a} \\
 p_i (1-S ) &=&-\frac{1}{2} \kappa t^2
\left(\rho_1+\rho_2-(P_1+P_2)\right), \label{2a}
\end{eqnarray}
where we have introduced
\begin{eqnarray}
S=p_1+p_2+p_3, \nonumber \\
Q=p^2_1+p^2_2+p^2_3,
\end{eqnarray}
and $i=1,2,3$.

Note that we obtain the same equations if we write the Einstein
equations in the form $R_{\mu \nu}=-\kappa (T_{\mu \nu}+T g_{\mu
\nu})$. It is easy to see that for the vacuum Kasner solution we
have $S=Q=1$.

\subsection{General solution for two-fluid Kasner cosmologies}

We are interested in studying cosmological scenarios filled with two
cosmic fluids which have barotropic equation of state
\begin{eqnarray}\label{EqS}
P_1(t)=\omega_1 \rho_1(t), \nonumber \\
P_2(t)=\omega_1 \rho_2(t),
\end{eqnarray}
with $\omega_1 $ and $\omega_2$ constants. Let us now consider the
direct integration of the field equations (\ref{1a}) and (\ref{2a}).
From three Eqs.(\ref{2a}) we conclude that, in order to satisfy all
these equations with self-consistent expressions for energy
densities and pressures $\rho_1$, $\rho_2$, $P_1$ and $P_2$, we must
impose the condition $S=1$. This implies that
$\rho_1+\rho_2=P_1+P_2$, and then
\begin{eqnarray*}
\rho_1+\rho_2=\omega_1 \rho_1+\omega_2 \rho_2.
\end{eqnarray*}
Thus by taking into account Eq.\ref{1a} we obtain for energy
densities
\begin{eqnarray}
\label{rho1} \kappa \rho_1=\frac{(1-Q)(1-\omega_2)}{2  (\omega_1-\omega_2) \, t^2}, \\
\label{rho2} \kappa \rho_2=\frac{(1-Q)(\omega_1-1)}{2
(\omega_1-\omega_2) \, t^2},
\end{eqnarray}
where $\omega_1 \neq \omega_2$. For $\omega_1 = \omega_2$ we obtain
the trivial case of a single fluid in a Kasner cosmology.

Since we are interested in a characterization of two-fluid Kasner
cosmologies, we shall consider that the weak energy condition (WEC)
holds and then we shall require for each source component that
$\rho_{_1} \geq 0$ and $\rho_{_2} \geq 0$. These conditions will
imply that
\begin{eqnarray}
\label{Q1} 0 \leq Q \leq 1,
\end{eqnarray}
for $\omega_2 \leq 1, \omega_1 \geq 1$ or $\omega_2 \geq 1, \omega_1
\leq 1$. Thus the condition~(\ref{Q1}) still holds as in the case of
a Kasner cosmology filled with a single ideal fluid.

Note that one of the fluids, say $\rho_1$, can satisfy the DEC (i.e.
$-1 \leq \omega_1 \leq 1$) or have a phantom behavior ($\omega_1 <
-1$), while another component, say $\rho_2$, has a state parameter
satisfying $\omega_2
> 1$, thus always violating DEC.

\subsection{Non-interacting cosmic fluids}

Usually the Universe is modeled with single perfect fluids and with
mixtures of non-interacting perfect fluids
\cite{Gunzig1,Gunzig4,Gunzig5,Gunzig2,Gunzig3}. This means that it
is assumed that there is no conversion (energy transfer) among the
components and that each of them evolves separately. Now in order to
study the following two cases for the superposition of two cosmic
fluids we shall use the conservation equation for Kasner
cosmologies, which is given by
\begin{eqnarray}\label{CE}
\dot{\rho}_1+\dot{\rho}_2+\frac{S}{t} (\rho_1+\rho_2+P_1+P_2)=0.
\end{eqnarray}
Thus for two non-interacting perfect fluids we can write the
standard conservation laws
\begin{eqnarray}
\label{q0} \dot{\rho}_1+\frac{S}{t} (\rho_1+P_1)=0, \nonumber \\
\dot{\rho}_2+\frac{S}{t} (\rho_2+P_2)=0,
\end{eqnarray}
which identically satisfy the general conservation
equation~(\ref{CE}). By taking into account the barotropic equations
of state~(\ref{EqS}) both of these equations can easily be
integrated obtaining
\begin{eqnarray}
\rho_i(t)=\frac{C_i}{t^{S(1+\omega_i)}},
\end{eqnarray}
with $i=1,2$. Now, since the total energy density $\rho$ scale as
$1/t^2$ we have that
\begin{eqnarray}
\frac{C_1}{t^{S(1+\omega_1)}}+\frac{C_2}{t^{S(1+\omega_2)}} \sim
t^{-2}.
\end{eqnarray}
This implies that the condition $S(1+\omega_1)=S(1+\omega_2)=2$ must
be imposed, leading finally to $\omega_1=\omega_2$. Thus both fluids
have the same state equation and then the non--interacting
superposition of two fluids is trivially equivalent to the standard
scenario with a single cosmic fluid.

\subsection{Interacting cosmic fluids}

In this case we can rewrite the conservation equation~(\ref{CE}) in
the following form:
\begin{eqnarray}
\label{q} \dot{\rho}_1+\frac{S}{t} (\rho_1+P_1) &=& q(t),\\
\label{-q} \dot{\rho}_2+\frac{S}{t} (\rho_2+P_2) &=& -q(t),
\end{eqnarray}
where is introduced a phenomenological coupling function $q(t)$.
Such a kind of interacting term has been considered before in the
literature \cite{ITQ4,ITQ2,ITQ3,ITQ1}. Note that if $q(t) > 0$ we
have that there exists a transfer of energy from the fluid $\rho_2$
to the fluid $\rho_1$. Again the general conservation equation
(\ref{CE}) is identically satisfied.

Now by considering barotropic equations of state (\ref{EqS}) we
conclude that the interaction term has the general form
$q(t)=q_0/t^3$, where $q_0$ is an arbitrary constant. Thus by
putting $q(t)=q_0/t^3$ into Eqs.(\ref{q}) and (\ref{-q}) we obtain
the following solutions for the energy densities:
\begin{eqnarray}\label{ED}
\rho_i(t)=\pm
\frac{C_i}{t^{S(1+\omega_i)}}+\frac{q_0}{(S(1+\omega_i)-2)t^2},
\end{eqnarray}
where $i=1,2$; $C_i$ are integration constants and the positive
(negative) sign corresponds to $i=1$ ($i=2$). Note that due to the
fact that in Kasner cosmologies the total energy density $\rho$
scales as $t^{-2}$, without any loss of generality, we can put
$C_i=0$ ($\omega_1 \neq \omega_2$).

Now, since the pressures are isotropic, from Eq.(\ref{2a}) we obtain
that $S=1$. Then by putting the expressions for energy densities
$\rho_1$ and $\rho_2$ from Eq.(\ref{ED}) (with $C_1=C_2=0$ and
$S=1$) into Eq.(\ref{1a}) we obtain
\begin{eqnarray}\label{formq}
q_0=\frac{(1-Q)(\omega_1-1)(1-\omega_2)}{2
\kappa(\omega_1-\omega_2)},
\end{eqnarray}
leading to the same expressions for energy densities (\ref{rho1})
and (\ref{rho2}) of Section II-A. It is interesting to note that the
interaction term is characterized by a coupling of the form
\begin{eqnarray}\label{qqq}
q(t)=(\omega_1-1) \theta \rho_1=(1-\omega_2) \theta \rho_2,
\end{eqnarray}
where $\theta$ is the expansion factor and in general for the metric
(\ref{KM}) is given by $\theta=S/t$. Notice that interacting
cosmological models with the interaction of the form (\ref{qqq})
were considered in the framework of flat FRW cosmologies
\cite{ICM1,ICM2,ICM3,ICM4,ICM5,ICM6}.

It is clear from conditions $\rho_1 \geq 0$ and $\rho_2 \geq 0$ that
we have the same constraint (\ref{Q1}) on $Q$. Thus from
Eq.(\ref{Q1}) and $\omega_2 \leq 1, \omega_1 \geq 1$ we obtain that
$q(t) \geq 0$ so we have energy transfer from $\rho_2$ to fluid
$\rho_1$; and from Eq.(\ref{Q1}) and $\omega_2 \geq 1, \omega_1 \leq
1$ we have that $q(t) \leq 0$ so we have energy transfer from
$\rho_1$ to the cosmic fluid $\rho_2$.

It calls to attention that the energy densities for the interacting
case coincide with the energy densities of the subsection II-A.
Mainly this is due to the power law character of scale factors in
the Kasner metric (\ref{KM}). In order to see this more clearly let
us discuss the general solution obtained in Section II-A. In
principle, on general expressions (\ref{rho1})--(\ref{rho2}) we can
now impose particular requirements. For example let us impose the
condition (\ref{q0}). Thus we obtain the following constraints on
state parameters: $\omega_1=\omega_2=1$. This implies that we really
have a single stiff fluid filling the Kasner cosmology. On the other
hand, if we now require that both cosmic fluids
(\ref{rho1})--(\ref{rho2}) interact with each other, then the
fulfillment of the conditions (\ref{q})--(\ref{-q}) does not add any
extra condition on the model parameters $\omega_1$, $\omega_2$ and
$Q$, defining only the final form of the interacting term
$q(t)=q_0/t^3$ (with $q_0$ expressed by Eq.(\ref{formq})). Thus the
form of the energy densities (\ref{rho1})--(\ref{rho2}) remains
unchanged. However, we can impose an explicit form on $q(t)$. For
example, for interacting terms given by $q(t)=\alpha \theta \rho_1$,
$q(t)=\beta \theta \rho_2$ or $q(t)=\gamma \theta \rho_1 + \delta
\theta \rho_2$ we shall obtain some constraints on $\omega_1$,
$\omega_2$, $Q$ with constant parameters $\alpha$, $\beta$ or
$\gamma$ and $\delta$ respectively.

\section{Three-fluid Kasner cosmologies}

In the following we shall consider Kasner cosmologies filled with
three barotropic cosmic fluids $\rho_1$, $\rho_2$ and $\rho_3$. It
is clear that the field equations now are given by
\begin{eqnarray}
\label{1atres} S-Q =\frac{1}{2} \kappa t^2
\left(\rho_1+\rho_2+\rho_3+3(P_1+P_2+P_3) \right),\\
\label{2atres} p_i (1-S )=-\frac{1}{2} \kappa t^2
\left(\rho_1+\rho_2+\rho_3-(P_1+P_2+P_3)\right),
\end{eqnarray}
where $i=1,2,3$. Now we shall consider a barotropic equation of
state $P_i=\omega_i \, \rho_i$ for each fluid. Since there are two
differential equations with three unknown functions $\rho_i$ , we
shall suppose $\rho_3$ as a given function, in order to close the
system of equations.

Requiring that $S=1$, from Eq.(\ref{2atres}) we obtain that
\begin{equation}\label{2atresss}
\rho_1 (1-\omega_1)+\rho_2 (1-\omega_2)+\rho_3 (1-\omega_3)=0.
\end{equation}
Solving Eqs.(\ref{1atres}) and (\ref{2atresss}) we obtain
\begin{eqnarray}
\label{rho31} \kappa \rho_1=\frac{(1-Q)(1-\omega_2)}{2
(\omega_1-\omega_2) \, t^2}+
\frac{\omega_2-\omega_3}{\omega_1-\omega_2} \kappa \rho_3, \\
\label{rho32} \kappa \rho_2=\frac{(1-Q)(\omega_1-1)}{2
(\omega_1-\omega_2) \,
t^2}+\frac{\omega_3-\omega_1}{\omega_1-\omega_2} \kappa \rho_3.
\end{eqnarray}
Let us now elucidate if all state parameters will respect DEC when
the fulfillment of WEC is required, i.e. $\rho_i \geq 0$ (for each
$i=1,2,3$). By adding Eqs.(\ref{rho31}) and (\ref{rho32}) and taking
into account condition $\rho_3 \geq 0$ we conclude that the
constrain (\ref{Q1}) is valid. In order to have different equations
of state for the cosmic ideal fluids we must require $\omega_1 \neq
\omega_2 \neq \omega_3$. This implies that, without any loss of
generality, we can put $\omega_3< \omega_2 <\omega_1$. Thus from
Eqs.(\ref{rho32}) and (\ref{Q1}) we conclude that we must require
that $\omega_1 > 1$ in order to fulfill the requirement $\rho_2 \geq
0$. Thus, always at least one of the three state parameters violates
DEC, implying that in three-fluid Kasner cosmologies it is not
possible to have simultaneously, for each fluid, $\omega_i < 1$ if
we require that $\rho_i \geq0$ ($i=1,2,3$). This leads to the
impossibility of having all three fluids simultaneously satisfying
DEC, i.e. $-1 \leq \omega_i \leq 1$.

Note that for this general solution we can invoke the conservation
equation, which is identically satisfied by energy densities
(\ref{rho31}) and  (\ref{rho32}), and has the following form:
\begin{eqnarray}
\label{a000} \dot{\rho}_1+\dot{\rho}_2+\dot{\rho}_3 \hspace{5.3cm} \nonumber \\
+\frac{1}{t} \left(\rho_1(1+\omega_1)+ \rho_2 (1+\omega_2)+\rho_3
(1+\omega_3)\right)=0.
\end{eqnarray}
Clearly, one can consider the case where a two-fluid configuration
is conserved separately from a third component, i.e. we have
\begin{eqnarray}
\label{a0} \dot{\rho}_1+\dot{\rho}_2+\frac{1}{t} \left(\rho_1(1+\omega_1)+ \rho_2 (1+\omega_2)\right)=0,\\
\label{c0} \dot{\rho}_3+\frac{1}{t} (1+\omega_3)=0.
\end{eqnarray}
Of course this is a particular solution of the above obtained
general solution. From Eq.(\ref{c0}) we obtain that
\begin{eqnarray}\label{rho30}
\rho_3(t)=C t^{-(1+\omega_3)},
\end{eqnarray}
where $C$ is an integration constant. Thus the solution in this case
takes the form
\begin{eqnarray}
\label{C2a} \kappa \rho_1=\frac{(1-Q)(1-\omega_2)}{2
(\omega_1-\omega_2) \, t^2}+ \frac{\kappa
C(\omega_2-\omega_3)}{(\omega_1-\omega_2) \, t^{1+\omega_3}},
\nonumber \\
\label{C2b} \kappa \rho_2=\frac{(1-Q)(\omega_1-1)}{2
(\omega_1-\omega_2) \, t^2}+\frac{\kappa
C(\omega_3-\omega_1)}{(\omega_1-\omega_2) \, t^{1+\omega_3}}.
\end{eqnarray}
Such a solution may describe multi-fluid Kasner cosmologies filled
with a two-fluid conserved configuration and independently conserved
dust ($\omega_3=0$), radiation ($\omega_3=1/3$), or even a
cosmological constant ($\omega_3=-1$), among others. If the third
fluid is a stiff one, i.e. $\omega_3=1$, this solution becomes a
scaling cosmological solution. Note that for the case where each
fluid component is conserved separately we obtain that
$\rho_3(t)=\frac{1-Q}{2 \kappa t^2}$ with $\rho_1=\rho_2=0$, so this
case is equivalent to the trivial scenario of a single stiff fluid
in a Kasner cosmology.

Now let us consider interacting scenarios for these three fluids
$\rho_1$, $\rho_2$ and $\rho_3$. We can thus write (where $S=1$ and
$P_i= \omega_i \rho_i$)
\begin{eqnarray}
\label{a} \dot{\rho}_1+\frac{\rho_1}{t} (1+\omega_1)&=&\alpha(t),\\
\label{b} \dot{\rho}_2+\frac{\rho_2}{t} (1+\omega_2)&=&\beta(t), \\
\label{c} \dot{\rho}_3+\frac{\rho_3}{t} (1+\omega_3)&=&\gamma(t),
\end{eqnarray}
where $\alpha$, $\beta$ and $\gamma$ must be constrained as follows:
\begin{eqnarray}\label{Cabc}
\alpha(t)+\beta(t)+\gamma(t)=0.
\end{eqnarray}
By introducing into Eqs.(\ref{a}) and (\ref{b}) the general
expressions for energy densities (\ref{rho31}) and (\ref{rho32}) we
obtain
\begin{eqnarray}
\label{aa} \alpha(t)&=&\frac{(1-Q)(\omega_1-1)(1-\omega_2)}{2
\kappa(\omega_1-\omega_2)t^3} \nonumber \\
&& +\frac{(1+\omega_1) (\omega_2-\omega_3)}{(\omega_1-\omega_2)t} \,
\rho_3(t)+
\frac{\omega_2-\omega_3}{\omega_1-\omega_2} \dot{\rho}_3, \\
\label{bb} \beta(t)&=&-\frac{(1-Q)(\omega_1-1)(1-\omega_2)}{2
\kappa(\omega_1-\omega_2)t^3}
 \nonumber \\
&& -\frac{(1+\omega_2)(\omega_1-\omega_3)}{(\omega_1-\omega_2)t}
\rho_3(t)+\frac{\omega_3-\omega_1}{\omega_1-\omega_2} \dot{\rho}_3.
\end{eqnarray}
Since Eqs. (\ref{rho31}) and (\ref{rho32}) are the general
expressions for energy densities in Kasner cosmologies filled with
relativistic ideal fluids $\rho_1$, $\rho_2$ and $\rho_3$, the
substitution of Eqs.(\ref{rho31}) and (\ref{rho32}) into
Eqs.(\ref{a})--(\ref{c}) yields the identical fulfillment of
constraint (\ref{Cabc}). Thus, in order to consider different
interacting scenarios we can impose a specific form on the
interacting term $\gamma(t)$, and then, by using Eqs.(\ref{aa}) and
(\ref{bb}), find interacting terms $\alpha(t)$ and $\beta(t)$ and
the final forms for energy densities (\ref{rho31}) and
(\ref{rho32}).

Let us for example consider the scenario where $\rho_1$ and $\rho_2$
are interacting with each other while $\rho_3$ is conserved
separately. Such a solution may describe a multi-fluid Kasner
cosmology filled with two interacting fluids and a conserved dust
($\omega_3=0$), radiation ($\omega_3=1/3$) or cosmological constant
($\omega_3=-1$), among others. This means that we must put
$\gamma(t)=0$, obtaining from Eq.~(\ref{c}) that the energy density
$\rho_3(t)$ is given by Eq.(\ref{rho30}). In this case
Eq.(\ref{Cabc}) becomes $\alpha+\beta=0$. For $C=0$ we obtain the
case of two interacting fluids discussed in Section II. Thus the
general solution in this case takes the form (\ref{C2b}) with
interaction terms given by
\begin{eqnarray}
\alpha(t)=\frac{(1-Q)(\omega_1-1)(1-\omega_2)}{2
\kappa(\omega_1-\omega_2)t^3} \nonumber \\
+\frac{C(\omega_1-\omega_3) (\omega_2-\omega_3)}{(\omega_1-\omega_2)
\, t^{2+\omega_3}} \, =-\beta(t).
\end{eqnarray}
Now, as another example, we shall consider a scenario where all
three fluids interact with each other. Let us suppose that
\begin{eqnarray}\label{gg}
\gamma(t)=\frac{q_{30}}{t^n},
\end{eqnarray}
where $q_{30}$ and $n$ are constants. Thus from Eq.(\ref{c}) we
obtain that the energy density of the third fluid is given by
\begin{eqnarray}
\rho_3=\frac{q_{30}}{(-n+2+\omega_3) \, t^{n-1}}+C
t^{-(1+\omega_3)},
\end{eqnarray}
which helps us to find the form of energy densities $\rho_1$ and
$\rho_2$ by Eqs.(\ref{rho31}) and (\ref{rho32}). Thus other
interacting terms are given by
\begin{eqnarray}
 \alpha(t)&=&\frac{(\omega_2-\omega_3)(\omega_1+2-n)
q_{30}}{(\omega_3-n+2)(\omega_1-\omega_2)t^n} \nonumber \\
&&+\frac{(1-Q)(\omega_1-1)(1-\omega_2)}{2
\kappa(\omega_1-\omega_2)t^3}\nonumber \\
&&+\frac{C(\omega_1-\omega_3)
(\omega_2-\omega_3)}{(\omega_1-\omega_2) \, t^{2+\omega_3}}, \nonumber \\ \nonumber \\
\beta(t)&=&\frac{(\omega_3-\omega_1)(\omega_2-n+2)
q_{30}}{(\omega_3-n+2)(\omega_1-\omega_2)t^3} \nonumber \\
&&-\frac{(1-Q)(\omega_1-1)(1-\omega_2)}{2
\kappa(\omega_1-\omega_2)t^3}\nonumber \\
&&-\frac{C(\omega_1-\omega_3)
(\omega_2-\omega_3)}{(\omega_1-\omega_2) \, t^{2+\omega_3}}.\nonumber \\
\end{eqnarray}
The interacting term given by Eq.(\ref{gg}) generalizes the kind of
interacting term discussed in Section II-C for which, to obtain it
we must put $n=3$. In this case, if additionally we have
$\omega_3=1$, this solution becomes a scaling cosmological solution.
For $q_{30}=0$ we obtain the previously discussed case of a
multi-fluid Kasner cosmology filled with two interacting fluids and
a single conserved one.

Since now we have three interacting terms, it is possible to have
two of them positive and one negative, or two negatives and one
positive. For example suppose that $\gamma <0$, $\alpha>0$ and
$\beta>0$. This implies that we have energy transfer from the third
fluid $\rho_3$ to the fluids $\rho_1$ and $\rho_2$. On the other
hand if $\gamma>0$, $\alpha<0$ and $\beta<0$ we have that energy is
transferred from fluids $\rho_1$ and $\rho_2$ to the cosmic fluid
$\rho_3$, so the fluids $\rho_1$ and $ \rho_2$ are being diluted due
to their interaction with $\rho_3$. Let us consider an explicit
example of such a scenario. If $n=3$, $q_{30}<0$, $C=0$, $\omega_3<
\omega_2<1$ (note that in this case $\rho_3>0$), $\omega_1>1$ and
$Q<1$ we have that $\gamma <0$, $\alpha>0$. For the other
interacting term we have that $\beta >0$ if $\tilde{q}<q_{30}<0$,
and $\beta<0$ if $q_{30} <\tilde{q}$, where
\begin{eqnarray*}
\tilde{q}=-\frac{(1-Q)(\omega_1-1)(1-\omega_3)}{(\omega_1-\omega_3)}.
\end{eqnarray*}
So we have that energy is transferred from $\rho_3$ to $\rho_1$ and
to $\rho_2$ if $\tilde{q}<q_{30}<0$; and for $q_{30} <\tilde{q}$,
fluids $\rho_3$ and $\rho_2$ transfer their energy to the cosmic
fluid $\rho_1$. It is interesting to note that this kind of triply
interacting fluid configurations also has been considered in the
framework of the FRW cosmologies \cite{Norman1,Norman2}.

\section{Conclusions}
In this paper we have provided a detailed analysis of Kasner
cosmologies dominated by two or three relativistic cosmic fluids.
Both cases can be exactly solved in the framework of Einstein field
equations. It was shown that the case, where each cosmic fluid
evolves separately according to standard conservation laws, leads us
to the trivial case of Kasner cosmologies dominated by a single
fluid; while if the anisotropic expansion is dominated by cosmic
fluids which are not conserved separately (both for two-fluid
configurations and at least two for three-fluid configurations),
then the cosmological scenarios are not at all trivial. For
two-fluid cosmological scenarios there exist only cosmological
scaling solutions. For three-fluid configurations, among
cosmological scaling solutions, there exist also more general ones.
It is shown that for two or three-fluid cosmological scenarios, by
requiring the positivity of energy densities, there always is a
matter component which violates DEC in this kind of anisotropic
cosmologies.

Finally, let us consider more precisely the general constraints
valid for multi-fluid Kasner cosmologies and expressed by
Eq.(\ref{Q1}) and $S=1$. Constraint (\ref{Q1}) implies that for each
$p_i$ we have that $-1<p_i<1$. By using $S=1$, we can replace
$p_3=1-p_1-p_2$ into $Q=p_1^2+p_2^2+p_3^2$. Now if we consider $Q$
as a given parameter we find that
\begin{eqnarray*}
p_1=\frac{1}{2}\left(1-p_2 \pm \sqrt{-3p_2^2+2p_2+2Q-1}\right),
\end{eqnarray*}
and then, in order to have a real $p_1$, we must require that
$-3p_2^2+2p_2+2Q-1 \geq0$. Thus the constraint $1/3 \leq Q \leq 1$
must be imposed. So the behavior of three scale factors
$a_i=t^{p_i}$ in multi-fluid Kasner cosmologies are restricted to
exhibiting decelerated expansions in all three directions, or even
contraction in several directions.

\acknowledgments

The authors thank Paul Minning for carefully reading this
manuscript. This work was supported by CONICYT through Grant
FONDECYT N$^0$ 1080530 and by PhD Grants N$^0$ 21070949 (FA) and
N$^0$ 21070462 (PM).

%\nocite{*}
%\bibliographystyle{spr-mp-nameyear-cnd}
%\bibliography{myref}

\begin{thebibliography}{}
\bibitem[Armendariz-Picon et al 2000]{Armendariz1} Armendariz-Picon, C.,
Mukhanov, V.F., Steinhardt, P.J., Phys. Rev. Lett. {\bf 85}, 4438
(2000).
\bibitem[Astier et al 2006]{Obs7} Astier, P. et al., Astron. Astrophys. {\bf 447}, 31
(2006).
\bibitem[Barrow 1997,1999]{Barrow} Barrow, J.D., Phys.\ Rev.\  D {\bf 55}, 7451
(1997), Barrow, J.D., Maartens, R., Phys.\ Rev.\  D {\bf 59}, 043502
(1999).
\bibitem[Barrow et al 2006]{ITQ3} Barrow, J.D., Clifton, T., Phys.\ Rev.\  D {\bf 73}, 103520
(2006).
\bibitem[Bean et al 2001]{Armendariz2} Bean, R.,  Magueijo, J., Phys. Lett. B {\bf 517}, 177 (2001).
\bibitem[Bennett et al 2003]{Obs4} Bennett, C. L. et al., Astrophys. J. Suppl 148, 1 (2003), astro-ph/0302225.
\bibitem[Bini et al 2007]{Others4} Bini, D., Cherubini, C., Jantzen, R.T.,
Class.\ Quant.\ Grav.\ {\bf 24}, 5627 (2007).
\bibitem[Bini et al 2009]{Bini} Bini, D., Cherubini, C., Geralico, A.,Jantzen, R.T.,
Class.\ Quant.\ Grav.\  {\bf 26}, 025012 (2009).
\bibitem[Bozza et al 2005]{Gunzig5} Bozza, V., Veneziano, G., JCAP {\bf 0509}, 007 (2005).
\bibitem[Brevik 1997]{Brevik} Brevik, I.H., Pettersen, S.V., Phys.\
Rev.\  D {\bf 56}, 3322 (1997)
\bibitem[Brevik et al 2000]{Cataldo152} Brevik, I.H., Pettersen, S.V., Phys.\ Rev.\  D {\bf 61}, 127305 (2000).
\bibitem[Cataldo et al 2000]{Cataldo151} Cataldo, M., del Campo, S.,
Phys.\ Rev.\  D {\bf 61}, 128301 (2000).
\bibitem[Cataldo et al 2001]{ScTT} Cataldo, M., del Campo, S., Salgado, P., Phys.\ Rev.\
D {\bf 63}, 063503 (2001).
\bibitem[Cataldo et al 2001]{HOLOGRAFIA} Cataldo, M., Cruz, N., del
Campo, S., Lepe, S., Phys.\ Lett.\  B {\bf 509}, 138 (2001).
\bibitem[Cataldo et al 2008]{Cataldos1} Cataldo, M., Mella, P., Minning, P., Saavedra, J.,
Phys.\ Lett.\  B {\bf 662}, 314 (2008).
\bibitem[Chimento et al 2003]{Cataldos4} Chimento, L.P., Jakubi, A.S., Pavon, D., Zimdahl, W., Phys. Rev. D {\bf 67} 083513, (2003).
\bibitem[Copeland et al 2010]{Others1} Copeland, E.J., Niz, G., Turok, N., Phys.\ Rev.\  D {\bf 81},
126006 (2010).
\bibitem[Cruz et al 2008]{Norman1} Cruz, N., Lepe, S., Pe\~na, F., Phys.\ Lett.\ B {\bf 663}, 338 (2008).
\bibitem[del Campo et al 2006]{ITQ4} del Campo, S., Herrera, R., Olivares, G., Pavon, D., Phys.\
Rev.\  D {\bf 74}, 023501 (2006).
\bibitem[Erickson et al 2004]{Turok} Erickson, J.K., Wesley, D.H., Steinhardt, P.J., Turok, N., Phys.\
Rev.\ D {\bf 69}, 063514 (2004).
\bibitem[Gavrilov et al 2004]{Gunzig3} Gavrilov, V.R.,  Melnikov, V.N.,  Abdyrakhmanov, S.T., Gen. Rel. Grav.{\bf 36}, 1579 (2004).
\bibitem[Goliath et al 2000]{Gunzig2} Goliath, M.,  Nilsson, U.S., J. Math. Phys. {\bf 41} 6906 (2000).
\bibitem[Gunzig et al 2000]{Gunzig1} Gunzig, E.,  Nesteruk, A.V.,  Stokley, M., Gen. Rel. Grav.{\bf 32}, 329 (2000).
\bibitem[Guo et al 2005]{Cataldos2} Guo, Z-K., Zhang, Y-Z., Phys. Rev. D {\bf 71}, 023501 (2005).
\bibitem[Halpern 2001]{Halpern} Halpern, P., Phys.\ Rev.\  D {\bf 63}, 024009
(2001).
\bibitem[Hinshaw et al 2006]{Obs2} Hinshaw, G. et al.,astro-ph/0603451.
\bibitem[Ivashchuk et al 2008]{Others3} Ivashchuk, V.D., Melnikov, V.N., Grav.\ Cosmol.\
{\bf 14}, 154 (2008).
\bibitem[Jamil et al 2008]{Norman2} Jamil, M., Rahaman, F., gr-qc/0810.1444.
\bibitem[Kofinas et al 2006]{ITQ1} Kofinas, G., Panotopoulos, G., Tomaras, T.N., JHEP
{\bf 0601}, 107 (2006).
\bibitem[Lee 2008]{BD} Lee, S., gr-qc/0811.1643.
\bibitem[Mak et al 2002]{Others6} Mak, M.K., Harko, T., Int.\ J.\ Mod.\ Phys.\  D {\bf 11}, 447 (2002).
\bibitem[Nojiri et al 2006]{ICM5} Nojiri, S., Odintsov, S.D., Phys.\ Lett.\  B {\bf 639}, 144 (2006).
\bibitem[Olivares et al 2008]{ICM1} Olivares, G., Atrio-Barandela, F., Pavon, D., Phys.
\ Rev.\  D {\bf 77}, 063513 (2008).
\bibitem[Padmanabhan 2005]{Padmanabhan} Padmanabhan, T., ``Understanding our universe:
Current status and open issues'' in 100 Years of
Relativity--Space--time Structure: Einstein and Beyond, Ashtekar,
A., (Editor), World Scientific (Singapore, 2005) pp 175-204;
arXiv:gr-qc/0503107.
\bibitem[Page et al 2006]{Obs3} Page, L. et al., astro-ph/0603450.
\bibitem[Pavon et al 2005]{Cataldos3} Pavon, D., Zimdahl, W., Phys.Lett. B {\bf 628}, 206 (2005).
\bibitem[Percival et al 2006,2007]{Obs5} astro-ph/0608636, Percival, W.J. et al., Astrophys. J. 657, 645
(2007).
\bibitem[Pinto-Neto 2005]{Gunzig4} Pinto-Neto, N., Santini, E.S., Falciano, F.T., Phys. Lett. A {\bf 344}, 131 (2005).
\bibitem[Ponce de Leon 2009]{Others2} Ponce de Leon, J., Class.\ Quant.\ Grav.\  {\bf 26},
185013 (2009).
\bibitem[Rashid et al 2009]{ICM6} Rashid, M.A., Farooq, M.U., Jamil, M., gr-qc/09013724.
\bibitem[Riess et al 2004]{Armendariz3} Riess, A. G. et al., Astrophys. J. {\bf 607}, 665
(2004).
\bibitem[Sadjadi e al 2006]{ICM4} Sadjadi, H.M., Alimohammadi, M., Phys.\ Rev.\  D {\bf 74}, 103007 (2006).
\bibitem[Setare et al 2009]{ICM3} Setare, M.R., Sadeghi, J., Amani, A.R., Phys.\ Lett.\  B {\bf 673}, 241 (2009).
\bibitem[Spergel et al 2003,2006]{Obs1} Spergel, D. N., et al., Astrophys. J. Suppl 148, 175 (2003),
astro-ph/0302209, Spergel, D.N. et al., astro-ph/0603449.
\bibitem[Srivastava 2006]{Armendariz4} Srivastava, S.K., Phys. Lett. B {\bf 643}, 1(2006).
\bibitem[Svitek 2006]{Others5} Svitek, O., Podolsky, J., Czech.\ J.\ Phys.\  {\bf 56}, 1367 (2006).
\bibitem[Tegmark et al 2004]{Obs6} Tegmark, M. et al., Phys. Rev. D {\bf 69},
103501 (2004).
\bibitem[Tomaras 2006]{ITQ2} Tomaras, T.N., hep-ph/0610412.
\bibitem[Zimdahl et al 2001]{ICM2} Zimdahl, W., Pavon, D., Phys.\ Lett.\  B {\bf 521}, 133 (2001).

\end{thebibliography}

\end{document}